# A validation methodology aid for improving a thermal building model : case of diffuse radiation accounting in a tropical climate

A.J.P. LAURET, T. A. MARA, H. BOYER, L. ADELARD, F. GARDE

*Université de La Réunion, Laboratoire de Génie Industriel, Equipe Génie Civil, BP 7151*
*15 avenue René Cassin, 97705 Saint-Denis Cedex, Ile de la Réunion, FRANCE*

**Abstract**

As part of our efforts to complete the software CODYRUN validation, we chose as test building a block of flats constructed in Reunion Island, which has a humid tropical climate. The sensitivity analysis allowed us to study the effects of both diffuse and direct solar radiation on our model of this building.
With regard to the choice and location of sensors, this stage of the study also led us to measure the solar radiation falling on the windows. The comparison of measured and predicted radiation clearly showed that our predictions over-estimated the incoming solar radiation, and we were able to trace the problem to the algorithm which calculates diffuse solar radiation. By calculating view factors between the windows and the associated shading devices, changes to the original program allowed us to improve the predictions, and so this article shows the importance of sensitivity analysis in this area of research.

## 1. Introduction

In recent years, major efforts have been made to develop methods for estimating the radiation flux falling on various surfaces (walls of buildings, solar and photo-voltaic panels, etc.). Many theoretical models are therefore available, covering both short and long wave radiation and also radiation sources (such as direct or diffuse radiation from the sky [1] or ground reflected diffuse radiation[2-3])

With regard to the interaction of these fluxes with the envelope of a building, their coupling with other transfer phenomena has also been studied intensively [4-5]. These aspects can be successfully incorporated into program codes, provided we take into account accepted (and justified) assumptions concerning the solar radiation (such as the partition between short and long wavelengths, the 'transparency' of the air as far as radiation are concerned, etc.) and also concerning the walls and windows (unidirectional conduction heat exchanges, constant radiation properties, etc.).

The calculation of the impact of near or further shading devices is also well-known, carried out using either trigonometry [6-7] or other methods [8]. Design of shading devices is easily accomplished by classical methods [9-10], but only in the case of direct beam radiation. However, most of the program codes do not calculate in detail the influence of near shading devices on diffuse radiation, apart from certain reference programs such as DOE-2 [11] and S3PAS [12]. In TRNSYS[13] for example, the shading reduction obtained for direct radiation is applied to the total incident radiation.

This article treats the solar radiation which enter a building through windows (or other glass frontages), and which are protected by a veranda, which is a very common set-up in tropical architecture. In this case, the simplifications mentioned earlier would lead us to overestimate the indoor solar radiation, and we will see that certain improvements should be made to the theoretical model.

## 2. The context of the study

*2-1 The Climate*

Reunion Island is situated at a latitude of 21° South and a longitude of 55° East. The climate is humid tropical. There is a dry season (May to October)- mainly cool and dry, predominated by the trade winds and a wet season (November to April)- hotter and wetter with light winds. In a humid tropical climate, the source of discomfort arises from a temperature increase due to bad architectural design, as far as insulation is concerned. Solar radiation accounts for 80% with the rest due to conduction exchanges. Under the humid tropical climate, the readings made on the short wave diffuse radiation show the importance of this kind of input. In particular, a quick analysis of the daily graphs on Reunion Island shows often significant diffuse radiation when the direct beam is low. The diffuse radiation therefore represents a significant fraction of the total (40% on the coast and about 50% for the interior of the island).

As a result, the solar protection of windows is fundamental, not only because this represents 15-20% of the thermal gains, but also because the windows contribute to the increase in the discomfort experienced by the occupant, due to the instant heating of the ambient air temperature and exposure to direct or reflected sunlight. All the windows must therefore be protected by some sort of shading device (overhanging horizontal canopies, screens, shadow masks etc.).

*2-1 The chosen building*

We have experimented with a typical dwelling of collective housing in Reunion island. This assessment is part of a technical evaluation of the building standards [14] for French overseas territories. The dwelling, represented in Fig. 1, includes three bedrooms and a living room. A veranda in effect enlarges the living room through the use of large bay windows.
Our analysis is based on the temperatures measured in the living-room, as the standard ECODOM concentrates exclusively on comfort improvement of living areas.

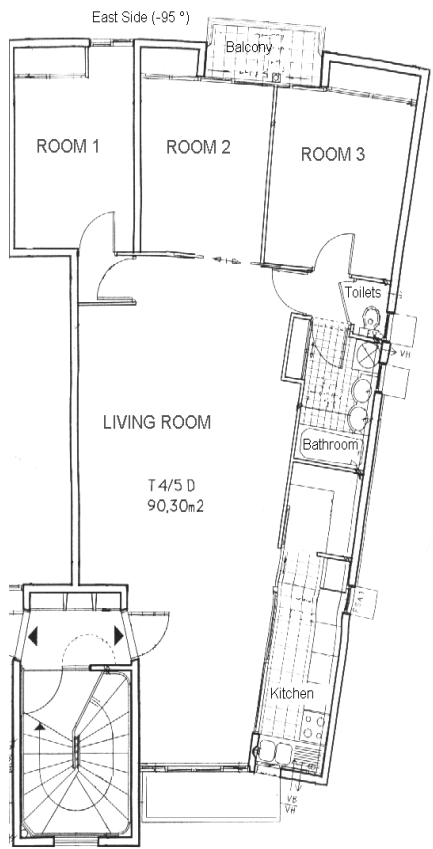

Fig.1. Typical dwelling

The measurement programme lasted throughout January 1998. For the meteorological data, equipment was set-up on the roof of the dwelling, giving half-hourly data for outdoor dry temperature, sky temperature, relative humidity, global and diffuse horizontal radiation and also wind speed and direction.

Ambient temperatures (dry and resultant*)* and relative humidity were measured in two places and at different heights in each room of the dwelling. Furthermore, some interior and exterior wall and roof surface temperatures were included in the experimental results.

*2-2 The software CODYRUN*

CODYRUN is a detailed building thermal simulation software regrouping design and research aspects. This software has already been covered in various publications [15-18]. Essentially, it involves multiple-zone software including natural ventilation and humidity transfer. Another key feature of CODYRUN is its multiple-model aspect. Indeed, before one simulation, the expert user has the possibility to choose between different models. The counterpart of this flexibility is the necessity for the building physics researcher to set up various models of different accuracy for each physical phenomenon.

The multiple-model aspect is detailed in [16], the thermal model in [17], the pressure airflow in [18], the data-structuration and the description of the front-end in [15]. Paper [14] focuses on both guidelines for tropical climate based on simulations with CODYRUN and split system modelling embedded in the software.

With regard to the validation of the program, CODYRUN was successfully 'BESTESTed' [21] and compared with other in situ measurements.

For the building chosen for this study, because of the large surface area of the bay windows (particularly on the West side), the largest energy source is from solar radiation. The performance of the building is therefore heavily influenced by the incoming radiation. It is recognised that for homes in cold climates with moderately sized windows and high levels of foundation insulation, energy simulation software should be fairly accurate [22]. The afore-mentioned article emphasises that for homes in hot climates, with large solar apertures and thermal mass, the situation is less clear. Simulation results for these situations should be used with caution.
Bearing in mind these elements, a recognised step-by-step validation procedure was used, including in particular a sensitivity analysis which is discussed in the following section.

# 3 Use of sensitivity analysis to plan the experiment

Sensitivity analysis (SA) of the model output is a very important stage in model building and analysis. In the building thermal simulation field, SA is being applied more and more [23-26]. Indeed, SA can help increase the reliability of the predictions made by building thermal simulation software.

This method identifies the most influential factors and inputs and evaluates their effect. In the validation methodology, the results are helpful as they give precise information on how to plan experiments. In particular, SA pinpoints the parameters or inputs that must be known accurately. In this way, we know which special measurements should be performed to ensure reliable predicted results.

*3.1. The methodology*

Using a MATLAB program which was specially developed for the chosen building, we conducted a sensitivity study, with the aim of determining the most significant parameters and input data for the prediction of the indoor dry air temperature. The analysis was complete, in the sense that it treated both parametric variables (material properties associated with the building envelope) and non-parametric (meteorological data). We will not give details of the method, as these can be found in another article [27].

The proposed method involves performing several simulation runs by oscillating each input sinusoidally over its range of interest. Analysing the spectrum (Fourier transform or power spectral density) of the output reveals the most influential factors.

1024 simulations were performed by making each factor (input or parameter) vary as a sinusoid ranging over ±10% with respect to its base-case value. In the following study, we are looking for the most important parameters for the predicted indoor air temperature ; once the simulations are achieved we can calculate the power spectrum density (PSD) of the residual.

As PSD is defined as the spectral decomposition of a signal's variance, we evaluate the influence of each parameter upon the output calculating the variance of the temperature differences ( $\text{var}(\Delta T_i)$ ) due to each parameter.

*3.2. Results*

The figure below shows that the input data and parameters which account for 90% of the residual variance are : direct and diffuse solar radiation, concrete walls thickness, density and specific heat capacity. These latter three parameters represent, in effect, the thermal inertia of the building. As far as the heat sources are concerned, this study clearly shows the importance of properly taking into account solar heat gains (both direct and diffuse).

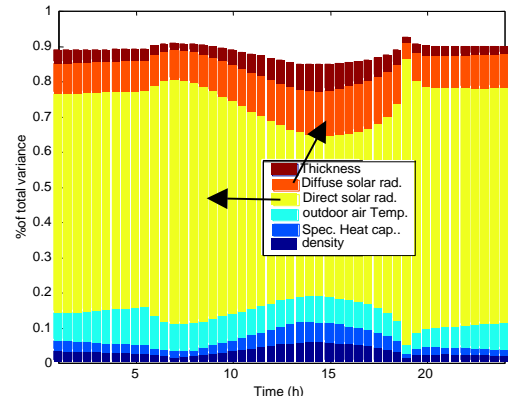

Fig.2. Assessment of the most important parameters

We note in particular that the contribution due to diffuse solar radiation increases at 3p.m.

As the absorptance of the walls for short wavelengths do not appear in the preceding analysis, the sensitivity of the apartment to short-wave radiation depends on solar gains through the windows.

Therefore, in order to determine the interaction between the building envelope and the outside solar radiation, in addition to the measurements mentioned earlier, we introduced two pyranometers on each side of the largest bay window. This was in the living room, facing West and so heavily exposed in the afternoon. In the photograph below, one can see the bay window as well as the shading screen. In this way, we obtained a measurement of the global incident solar radiation (without the shading effect, because this was placed at the edge of the plant tub) and of that transmitted by the bay (after the effect of the shading device and the glass).

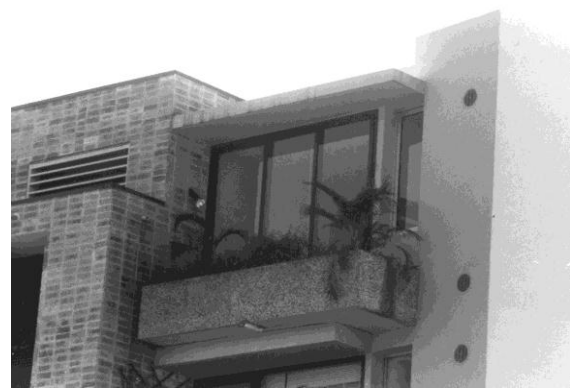

Fig.3. The dwelling (West side)

The initial detailed simulation software (CODYRUN) was modified to include the exterior global and interior radiation calculations (for each window) in the output file.

## 4. Initial comparison of model and measurements

For this 'initial' comparison, the diffuse exterior are considered to be unaffected by the presence of near-field shading. For the weather conditions experienced at the time of the measurements, all the exterior and interior openings were closed and cautiously sealed. The building was therefore behaving in a thermally pure mode, with no incoming airflow. For modelling purposes, five zones were considered, including the zone 'living room'. All our analysis is centred on comparisons of measurements with the model for dry air temperatures in the living room.

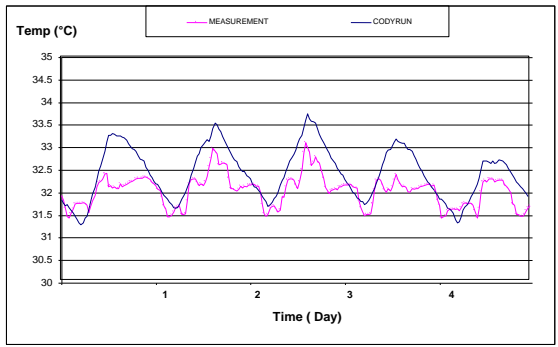

Fig.4. Dry air temperature of the living room

Figures 4 and 5 show the comparison between the predicted indoor dry air temperature in the living room and its measurement. The maximum acceptable measurement error is $\pm 0.5°C$. As the residual does not fall within these limits, the model is rejected. The mean and standard deviation are the following :

| Mean | Std |
|---|---|
| -0.39 | 0.369 |

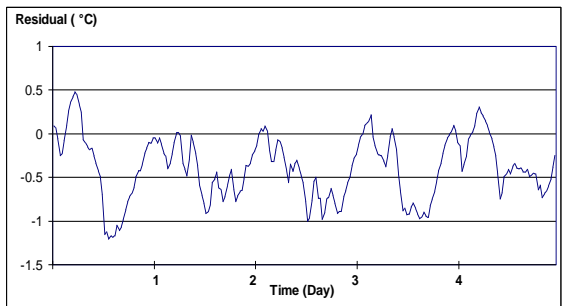

Fig.5. Residual

To find the source of the error, one has to search for correlation between the inputs and the residual. This study may guide the building physics researcher in the right direction for improving the model.

### 4.1. Use of time-frequency analysis

Instead of classical spectral analysis [28-29] that shows some limitations when the inputs are correlated (as in the case of meteorological data), time-frequency analysis is a powerful tool capable of de-segregating the actual effects of several inputs at different time-scales.

Time-frequency analysis is a modern technique to extract information from signals. The first and natural time-frequency tool is the short-scale Fourier Transform (SFTT). The key feature of these tools is that they simultaneously decompose a signal in time and frequency. One can imagine that the information obtained is richer than that given in only spectral or in time domain analysis. The representation of an SFTT is called the spectrogram. The goal of our study is to check whether the spectrogram of the residual is similar to the one of the inputs.

In order to obtain useful spectograms, low frequency data must be filtered out. For this purpose, a fourth-order Butterworth is used to remove frequencies lower than $0.02h^{-1}$. Filtering is justified in our case as the range of interest is around 1 day$^{-1}$ harmonic (~$0.041h^{-1}$).

Unlike classical spectral analysis, spectograms of climatic data show that the time-frequency signature of the outdoor air-temperature is very different from that of the solar radiation. It is possible here to distinguish outdoor air temperature from the other inputs. The main drawback of classical frequency analysis is avoided.

*4.2. Application to our case*

Spectograms of diffuse and direct solar radiation illustrate the complementarity of these two inputs. Consequently, it will be difficult to identify which of these 2 inputs is really correlated with the residual. Nevertheless, physically, the influence of the diffuse component seems more relevant. Comparison of the 2 spectograms clearly shows a correlation between diffuse solar radiation and the residual at times 90 and 130. The source of the error seems to be in the way the model account for diffuse solar radiation

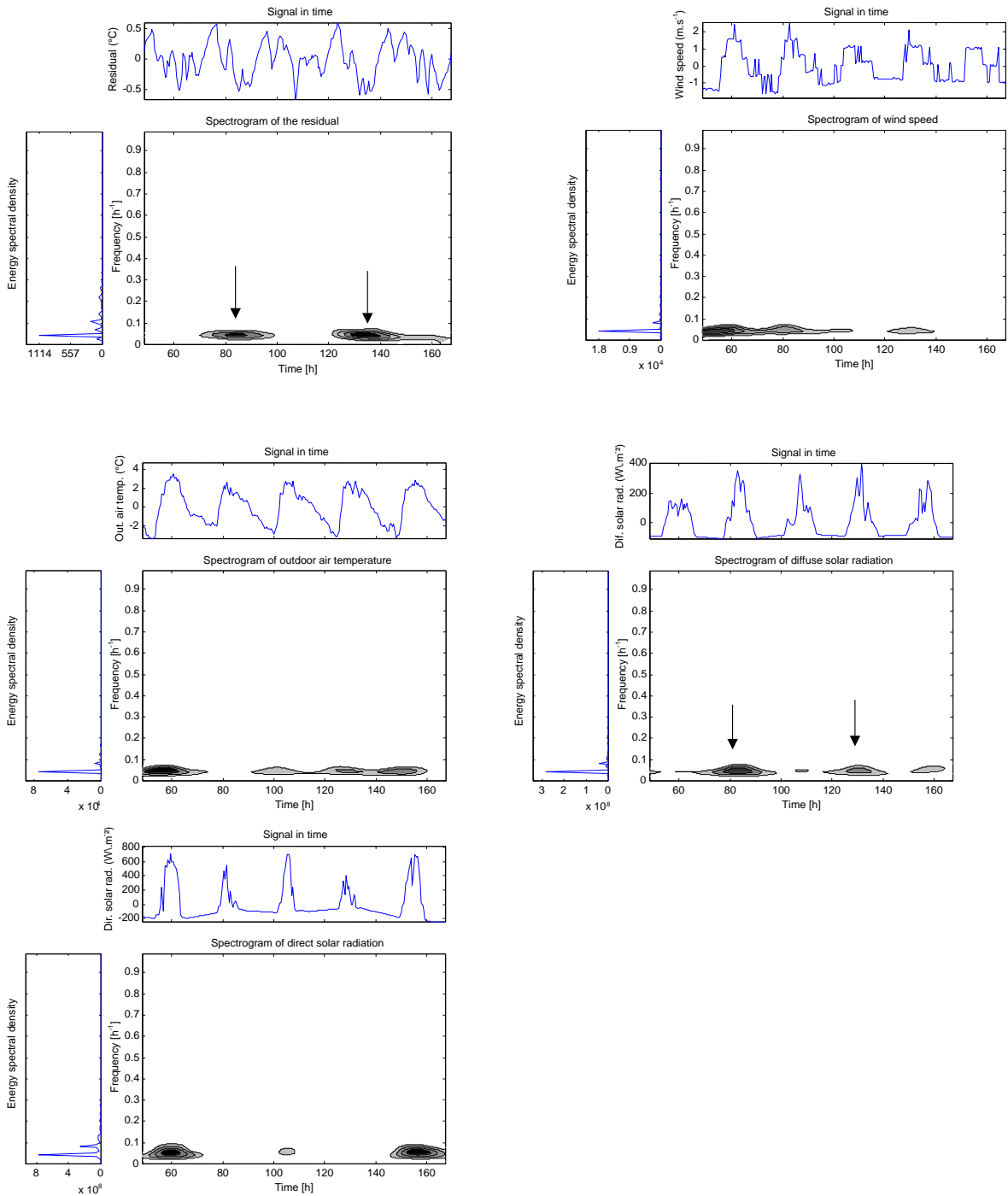

Fig.6. Spectograms of the residual and the inputs

## 5. Improvement of the model

*5.1. Comparison of incoming radiation*

Because of the two pyranometers installed after the sensitivity analysis, we are able to compare the global exterior solar radiation with the interior one, both for the actual measurements and for the model predictions. Given that the agreement for the exterior values is excellent, we clearly see that our initial model overestimates by 30 to 40% the interior radiation contributions in the living room (see fig.7).

Considering that the measurement error is approximately 6%, and that the error due to imprecise positioning of the pyranometers is neglectable, an error of 30 to 40% is evidently very significant. Further, as the sensitivity analysis had shown that the transmittance of the glass was not an important factor, the near shading device formed by the protection around the west bay window is clearly responsible for the overestimation of the solar gains. As CODYRUN has already been tested with respect to the effect of shading devices on the direct radiation [21], we decided to improve the model by introducing the possibility of reducing the diffuse solar radiation with near shading devices.

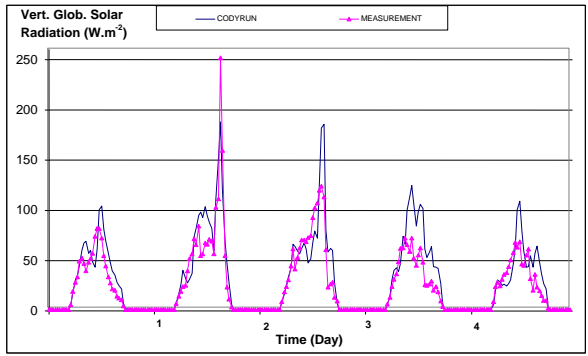

Fig.7. Comparison of the global solar radiation

*5.2 The chosen method*

As a near shading device is generally composed of several elements (top-flap, vertical fins, low walls,…), the calculation of the analytic shape factor involving the window and the shading device proved complex. We therefore preferred a simplified approach which uses properties of the view factors. By using projections perpendicular to the planes containing the windows and each element of the shade, the calculation of the global shape factor can be achieved by combining several view factors of the same type ; these view factors are those obtained for orthogonal parallelepipeds with a common edge [29]

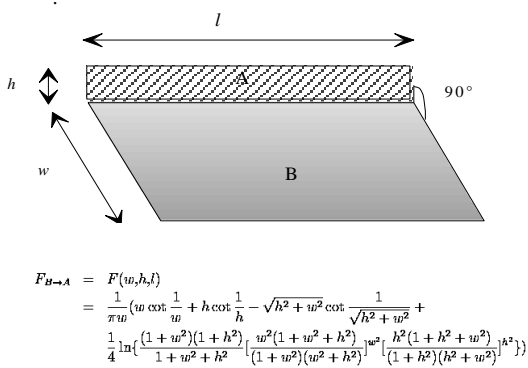

Fig.8. Calculation of the shape factor

For the West bay window which concerns us here (see Fig. 3), this approach requires us to find, in the following figure 9, the view factor between the window (coloured grey) and the elements composing the shade (hatched).

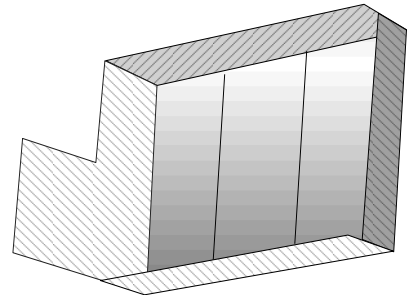

Fig.9. The bay window in its egg-crate

As a first step, we limited ourselves to one dimension per element of the shade. In this way the left side was represented by a parallelepiped of width equal to the top side. An alternative, more comprehensive approach has been tried, but was not integrated into the program, because of the small (approximately 1%) reduction obtained.

One should note that our method does not take into account the solar radiation reflected off the shade. This has been studied by Barozzi [6] for example, and could prove useful for a more detailed calculation of the radiation diffused around the window. Finally, the calculation for a near shading device uses a constant percentage of reduction for the diffuse solar radiation.

*5.3. Integration into CODYRUN*

Figure 10 describes the data inputs associated to the West bay window. The dimensions of the elements composing the shade are given in a standard way. Some parts of the data input window are disabled because some of these dimensions are considered infinite (the width of the top flap, for example).

Given the importance of the effect of the near shading devices on the incident diffuse around the window (see the preceding method), the percentage of reduction of diffuse is now indicated and taken into account at each step of simulation.

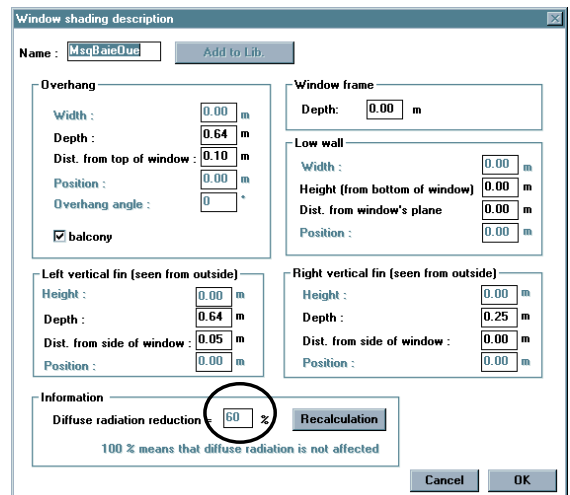

Fig. 10. *Shadow mask data capture window*

Buildenv12

In the case which interests us, 60% of the incident diffuse radiation on the West wall are blocked by the near shading device before reaching the glass.

## 6. New comparison

*6.1. Comparison of fluxes*

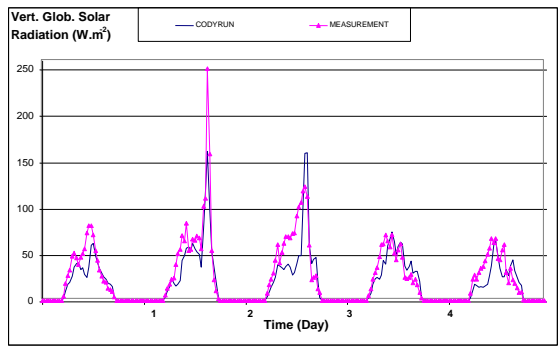

Fig.11. New comparison of the global solar radiation

In the interior, the new model is manifestly better than the old one at calculating the radiation transmitted into the zone. The relative average daily error in the density of transmitted flux has changed from 22% to 4% over the whole period. In particular, the prediction due to the model is much improved during days 4 and 5, which correspond to the periods when the diffuse radiation was dominant (the relative error is reduced from 32% to 1% in days 4 and 5).

It should be noted that the model used here to represent the diffuse radiation is known as 'isotropic'. Other simulations have shown us that, as could be expected, the choice of a non-isotropic model (in which some of the diffuse radiation is directional) modifies slightly the incoming energy contributions. In fact, the direct and diffuse parts of the flux falling on the bay window are respectively reduced by the direct solar factor associated with the shading device and the reduction factor for the diffuse part. Our building model is therefore sensitive to the diffuse radiation reconstitution.

### 6.2 **Comparison of temperatures**

The variable which interested us most was the dry air temperature in the living room, and the modification of the model led to a better prediction of this temperature (fig. 12). Thus, over the 5 day sequence considered, the average residual changed from about 0.4°C to 0°C, with a standard deviation of about 0.3°C slightly improved (from 0.37 to 0.34). Further, by concentrating the study on one day when the diffuse solar radiation was more important, an even better improvement was observed.

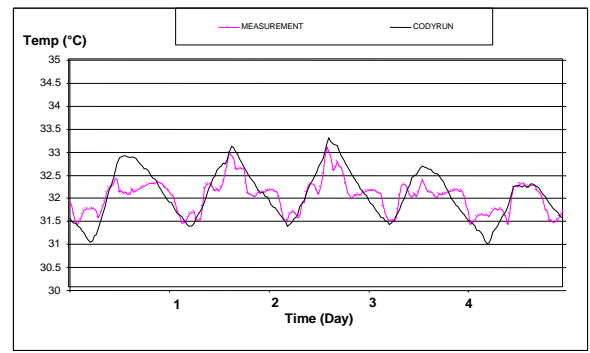

Fig.12. New comparison of the dry air temperature

Regarding the validation, the new model has allowed us to increase the number of points for which the residual is included within the maximum acceptable error limits, that is -0.5 to +0.5 °C.

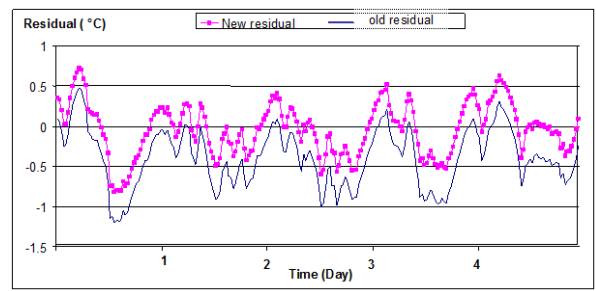

Fig.13. New and old residual

As the residual curve obtained was nearly always within these limits, the new building model is therefore considered as improved.

| Mean | Std |
|---|---|
| -0.04 | 0.336 |

We should note however that the method only leads to an improvement in the prediction of the steady state (correction of the mean). It is nevertheless an important step, which has to be completed before moving on to improve the prediction of the dynamic response.

## 7. Conclusion

We have used a step-by-step approach to validate the thermal behaviour of a simulation program. We have seen the advantages of a systematic application of the validation procedure. The sensitivity analysis combined with appropriate sensors has allowed us to highlight the weaknesses of the model for calculating the energy contributions due to diffuse solar radiation from the windows. In the case of experimental validation, sensitivity analysis has once again proved its ability to be a powerful tool for improving simulation models.

Regarding the aspects as basic as the calculation of the solar gains due to the windows, it is noteworthy that the use of computer codes written for an architecturally and climatically different context should be done with care. In conclusion, this paper illustrates that under our tropical climate and for the architectural rules currently applied, an adapted model must be used.